\documentclass{iau-JDSS}
\usepackage{graphicx}

\title[] %% give here short title %%
{Numerical Modeling of Multiphase, Turbulent Galactic Disks with Star
Formation Feedback}

\author[]   %% give here short author list %%
{Chang-Goo Kim$^{1}$, Eve C. Ostriker$^{2,3}$ 
 \break \and Woong-Tae Kim$^{4}$}

\affiliation{$^1$Department of Physics \& Astronomy, Western University, Canada \break email: ckim256@uwo.ca\\[\affilskip]
$^2$Department of Astronomy, University of Maryland, USA\\[\affilskip] 
$^3$Astrophysical Sciences, Princeton University, USA\\[\affilskip] 
$^4$Department of Physics \& Astronomy, Seoul National University, Republic of Korea\\[\affilskip]
}

\pubyear{2012}
\volume{Volume 16}  %% insert here IAU Highlights of Astronomy Volume Number
\pagerange{}
\date{?? and in revised form ??}
\jname{Highlights of Astronomy, Volume 16}
\editors{Thierry Montmerle, ed.}

\begin{document}

\maketitle

\begin{abstract}
Star formation is self-regulated by its feedback that drives turbulence and
heats the gas. In equilibrium, the star formation rate (SFR) should be directly
related to the total (thermal \emph{plus} turbulent) midplane pressure and
hence the total weight of the diffuse gas if energy balance and vertical
dynamical equilibrium hold simultaneously. To investigate this quantitatively,
we utilize numerical
hydrodynamic simulations focused on outer-disk regions where diffuse atomic gas
dominates. By analyzing gas properties at saturation, we obtain relationships
between the turbulence driving and dissipation rates, heating and cooling
rates, the total midplane pressure and the total weight of gas, and the SFR and
the total midplane pressure. We find a nearly linear relationship between the
SFR and the midplane pressure consistent with the theoretical prediction.

\keywords{stars: formation, ISM: kinematics and dynamics, methods: numerical, turbulence}

\end{abstract}


Star formation feedback is a key ingredient to model the star formation rates
(SFR) in galactic disks. The interstellar medium (ISM), the raw material for
star formation, is highly dissipative so that within a timescale comparable to
the dynamical timescales, such as the gravitational free-fall time and the
vertical oscillation period, the thermal energy is radiated away, and the
turbulent energy is dissipated by shocks and nonlinear cascades. Since the gas
depletion time, the time required to convert all the gas into stars, is much
longer than the dynamical timescales (e.g., \cite{kru12}), there must be
continuous and efficient mechanisms for energy replenishment. Feedback from
massive stars is the most probable source that provides a substantial amount of
momentum via expanding supernova remnants and photoelectric heating by FUV
radiation.

The turbulent and thermal pressures maintained by energy balance support
galactic disks against the total weight of the gas under the gravity of gas,
stars, and dark matter.  In equilibrium, the SFR is thus self-regulated to
supply the appropriate amount of thermal and turbulent energy that meets the
needs of the ISM.  Recently, \cite{OML10} and \cite{OS11} have developed an
analytic theory for regulation of SFRs based on the equilibrium model, which
successfully explains observed relationships among the SFR surface density, the
total and molecular gas surface densities, and the stellar surface density.

To directly test the analytic theory quantitatively, we utilize a series of
numerical hydrodynamic simulations that resolve vertical dynamics of the ISM
and include self-gravity, cooling and heating, and star formation feedback (see
also \cite{KKO11}). In our models, star formation feedback is realized by
time-dependent heating and by expanding supernova remnants.  We focus on outer
disk regions where the diffuse ISM dominates, with gas surface densities
$\Sigma =3-20 {\rm\;M_{\rm \odot}\;pc^{-2}}$ and star-plus-dark matter volume
densities $\rho_{\rm sd}=0.003-0.5  {\rm\;M_{\rm \odot}\;pc^{-3}}$.

\begin{figure}
\includegraphics[width=0.32\textwidth]{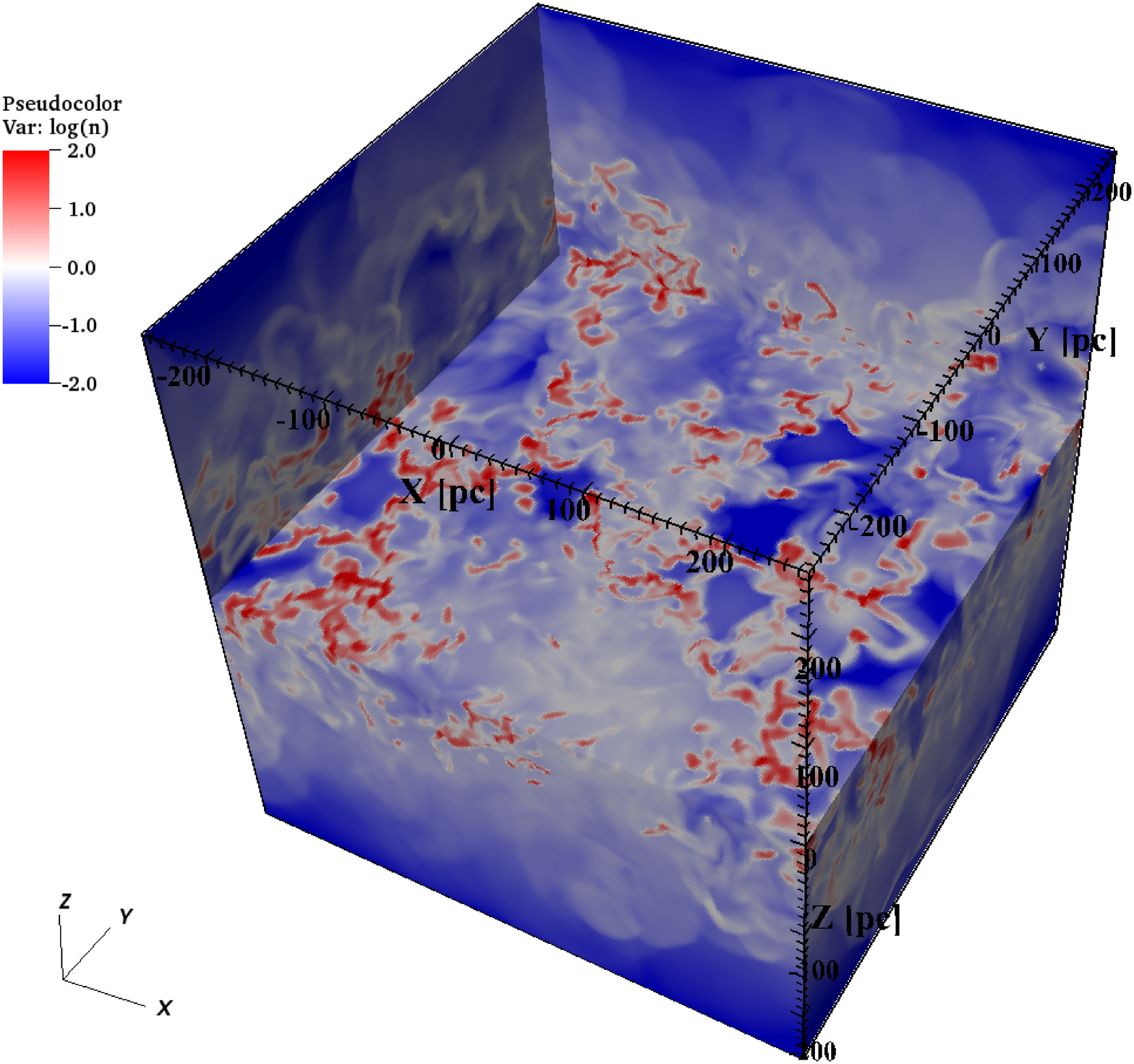}
\includegraphics[width=0.37\textwidth]{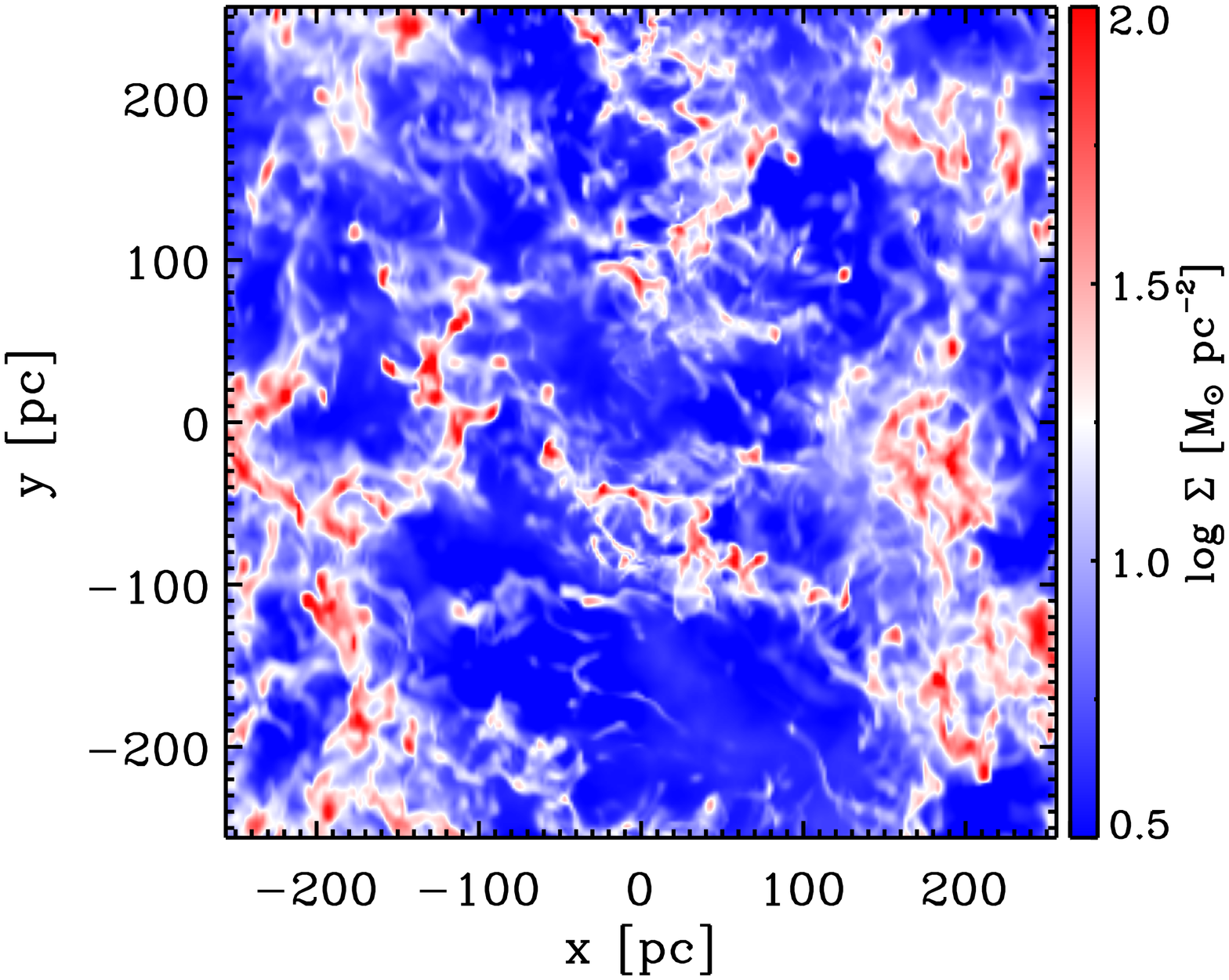}
\includegraphics[width=0.29\textwidth]{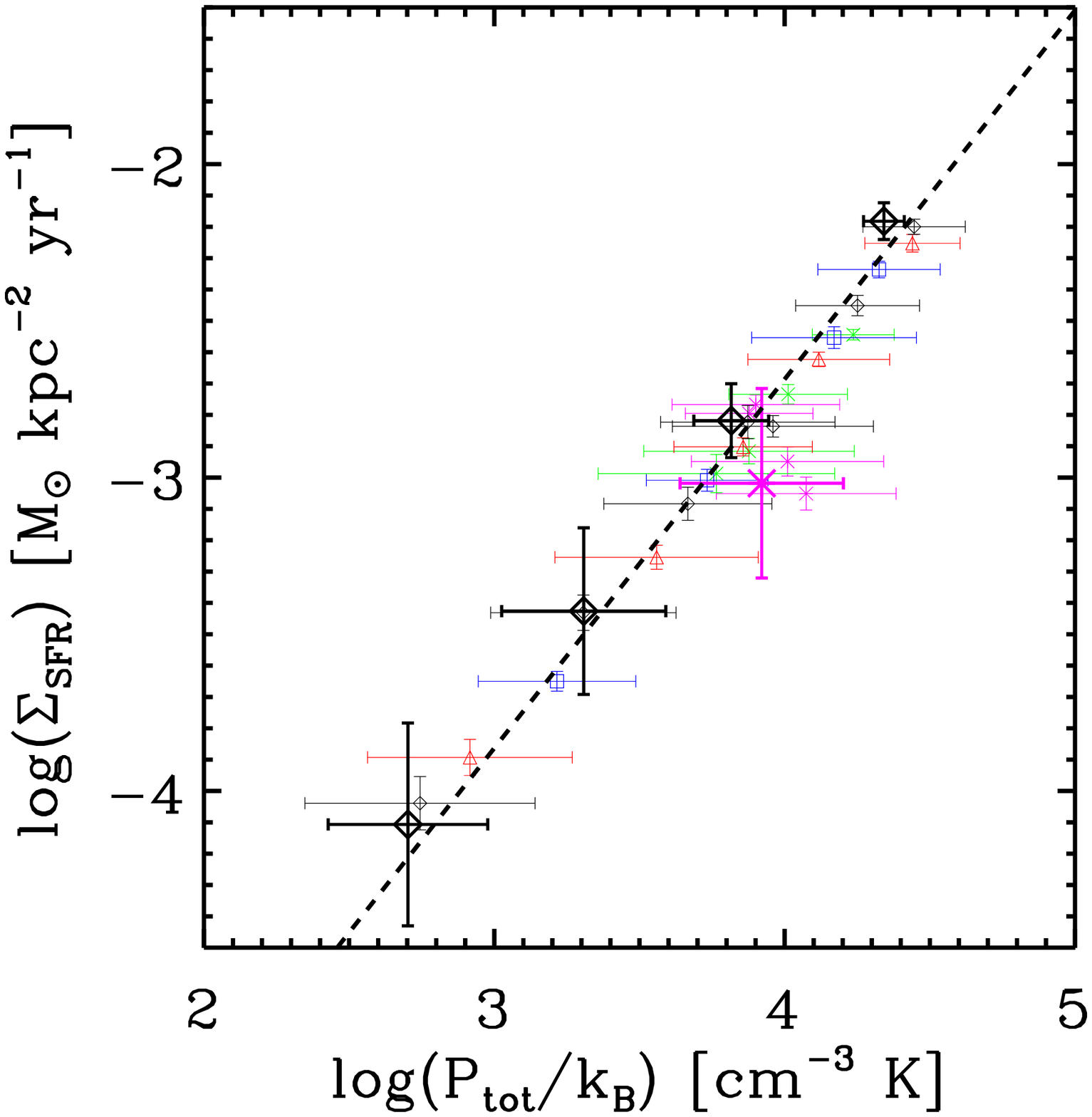}
\caption{
\emph{Left} and \emph{Middle}: Saturated state snapshots of the fiducial model
with $\Sigma=10 {\rm\;M_\odot\;pc^{-2}}$ and $\rho_{\rm sd}= 0.05
{\rm\;M_\odot\;pc^{-3}}$. The volume density slices (\emph{left}) and the
surface density seen along the vertical direction (\emph{middle}) are drawn in
logarithmic color scales.  \emph{Right}: SFR surface density as a function of
the total midplane pressure at a saturated state for all models. The dashed
line is our best fit with a slope of $1.18$.  }\label{fig}
\end{figure}

Our model disks undergo a quasi-periodic cycle of vertical oscillations: the
disk expands vertically due to feedback, reducing the SFR, which in turn causes
the disk to contract back, increasing the SFR and feedback. After one or two
vertical oscillations of the disk, the overall physical properties are fully
saturated.  Figure~\ref{fig} shows a morphology of the gas in the fiducial
model with $\Sigma=10 {\rm\;M_\odot\;pc^{-2}}$ and $\rho_{\rm sd}= 0.05
{\rm\;M_\odot\;pc^{-3}}$ at saturation in the left and middle panels; the
ISM is multiphase, turbulent, and highly-structured, which is qualitatively
similar to that seen in recent HI observations (see McClure-Griffith's
contribution in this volume).  We directly measure for all models the midplane
thermal ($P_{\rm th}$) and turbulent ($P_{\rm turb}$) pressures as well as the
SFR surface density ($\Sigma_{\rm SFR}$) that are averaged over one orbital
period corresponding typically to three or four vertical oscillations. The
thermal and turbulent pressures are approximately linearly proportional to the
SFR surface density as the equilibrium model predicts; $P_{\rm th}\propto
\Sigma_{\rm SFR}^{0.86}$, and $P_{\rm turb}\propto \Sigma_{\rm SFR}^{0.89}$. At
the same time, the total (thermal \emph{plus} turbulent) midplane pressure
($P_{\rm tot}$) is in excellent agreement with the dynamical equilibrium
pressure, resulting in $P_{\rm tot}\propto \Sigma\sqrt{\rho_{\rm sd}}$.
Finally, the fundamental relationship between the SFR surface density and the
total pressure naturally reflects the match between supply and demand of the
ISM, $\Sigma_{\rm SFR} \propto P_{\rm tot}^{1.18}$ (see the right panel of
Figure~\ref{fig}).

\begin{acknowledgments}
This work was made possible by the facilities of the Shared Hierarchical
Academic Research Computing Network (SHARCNET:www.sharcnet.ca) and
Compute/Calcul Canada. C.-G. K. is supported by a CITA National Fellowship.
\end{acknowledgments}


\begin{thebibliography}{}
%\bibitem[Field \etal\ (1969)]{fie69}
%	{{Field}, G.~B., {Goldsmith}, D.~W., \& {Habing}, H.~J.} 1969, 
%	\textit{ApJ} 155, L149

%\bibitem[Kim \etal\ (2010)]{KKO10}
%	{{Kim}, C.-G., {Kim}, W.-T., \& {Ostriker}, E.~C.} 2010, 
%	\textit{ApJ} 720, 1454

\bibitem[Kim \etal\ (2011)]{KKO11}
	{{Kim}, C.-G., {Kim}, W.-T., \& {Ostriker}, E.~C.} 2011, 
	\textit{ApJ} 743, 25

\bibitem[Krumholz \etal\ (2012)]{kru12}
	{{Krumholz}, M.~R., {Dekel}, A., \& {McKee}, C.~F.} 2012, 
	\textit{ApJ} 745, 69

%\bibitem[Krumholz \& Tan (2007)]{kru07}
%	{{Krumholz}, M.~R., \& {Tan}, J.~C.} 2007,
%	\textit{ApJ} 654, 304

%\bibitem[Stone \etal\ (1998)]{sto98}
%	{{Stone}, J.~M., {Ostriker}, E.~C., \& {Gammie}, C.~F.} 1998, 
%	\textit{ApJ} 508, L99

%\bibitem[Mac Low \etal\ (1998)]{mac98}
%	{{Mac Low}, M., {Klessen}, R.~S., {Burkert}, A., \& {Smith}, M.~D.} 1998,
%	\textit{Phys. Rev. Lett.} 80, 2754

\bibitem[Ostriker \etal\ (2010)]{OML10}
	{Ostriker, E.~C., McKee, C.~F., \& Leroy, A.~K.} 2010, 
	\textit{ApJ} 721, 975

\bibitem[Ostriker \& Shetty (2011)]{OS11}
	{Ostriker, E.~C., \& Shetty, R.} 2011, 
	\textit{ApJ} 731, 41 

%\bibitem[Kim \etal\ (2006)]{KKO06}
%	{{Kim}, C.-G., {Kim}, W.-T., \& {Ostriker}, E.~C.} 2006, 
%	\textit{ApJ} 649, L13

%\bibitem[Shetty \& Ostriker (2012)]{SO12} 
%	{Shetty, R., \& Ostriker, E.~C.} 2012,
%	\textit{ApJ} 754, 2

\bibitem[Wolfire \etal\ (1995)]{wol95}
	{{Wolfire}, M.~G., {Hollenbach}, D., {McKee}, C.~F., 
	{Tielens}, A.~G.~G.~M., \& {Bakes}, E.~L.~O.} 1995
	\textit{ApJ} 443, 152

\end{thebibliography}
\end{document}